\documentclass[12pt]{iopart}

\usepackage{iopams}
\usepackage{graphicx}

\begin{document}

\title[The spectrum of powers of the Laplacian in bounded domains]{The spectrum of large powers of the Laplacian in bounded domains}

\author{E Katzav and M Adda-Bedia}

\address{Laboratoire de Physique Statistique de l'Ecole Normale Sup\'erieure, CNRS UMR 8550, 24 rue Lhomond, 75231 Paris Cedex 05, France.}
\eads{\mailto{eytan.katzav@lps.ens.fr}, \mailto{adda@lps.ens.fr}}

\begin{abstract}

We present exact results for the spectrum of the N$^{th}$ power of the Laplacian in a bounded domain. We begin with the one dimensional case and show that the whole spectrum can be obtained in the limit of large $N$. We also show that it is a useful numerical approach valid for any $N$. Finally, we discuss implications of this work and present its possible extensions for non integer $N$ and for $3D$ Laplacian problems.

\end{abstract}

\vspace{2pc}

\submitto{\JPA (\today)}
\date
\maketitle

\section{Introduction and main results}

Recently there has been a growing interest in the problem of finding the spectrum of large powers of the Laplacian in bounded domains \cite{B&Ha,B&Hb,Pomeau}. In one dimension, which is the case that will be presented in detail here, the problem is simply that of finding eigenfunctions and eigenvalues to the equation
\begin{equation}
(-\Delta) ^N u\left( x \right) = \lambda u\left( x \right) \quad
\quad \quad \quad x \in \left[-1,1\right]
 \label{eq:eigen} \, ,
\end{equation}
for functions obeying the following Boundary Conditions (BC)
\begin{equation}
u\left( { \pm 1} \right) = u^{\left( 1 \right)} \left( { \pm 1}
\right) =  \cdots  = u^{\left( {N - 1} \right)} \left( { \pm 1}
\right) = 0
 \label{eq:BC} \, ,
\end{equation}
where $u^{(k)}(x)$ is the $k^{th}$ derivative of $u(x)$. Due to the invariance of (\ref{eq:eigen},\ref{eq:BC}) under the symmetry $x \rightarrow -x$, the eigenfunctions are characterized by a definite parity, i.e. $u(x)=\pm u(-x)$. In order to simplify the discussion it is useful to consider
separately the eigensystem of (\ref{eq:eigen},\ref{eq:BC}) for the even and odd eigenfunctions, which we will call from now on the even/odd spectrum.

From a mathematical point of view, Ref.~\cite{B&Ha} has shown that the determination of the spectrum of $\Delta^N$ can be related to four different problems - the spectrum of certain positive definite Toeplitz Matrices, the norm of the Green kernels of $\Delta^N$, the best constants in certain Wirtinger-Sobolev inequalities, and the conditioning of a special least squares problem. From a physical point of view, the interest in the spectrum of $\Delta^N$ comes from many directions. Classical problems such as diffusion and wave propagation require knowledge of the spectrum of $\Delta$ (i.e., $N=1$), which is also related to recent problems such as diffusion limited aggregation and chaos. Problems in elasticity theory often deal with $\Delta^2$ \cite{Muskhelishvili}, and the fractional Laplacian (i.e., when $N$ is not necessarily an integer) appears naturally in stochastic interfaces and L\'evy flights \cite{Zoia,fractional} and so knowledge of the spectrum allows progress in the understanding of anomalous diffusion and first-passage problems of a L\'evy flyer.

The case $N=1$, emanating from classical diffusion problems, is trivial. The even spectrum is composed of the set of eigenvalues $\left\{\left[\pi(j+1/2)\right]^2\right\}_{j=0,1,...}$ and their corresponding eigenfunctions $\cos(\pi(j+1/2)x)$ (or $(\pi j)^2$ and $\sin(\pi j x)$ respectively, for the odd spectrum). The case $N=2$, which is encountered when analyzing the vibrational modes of a clamped rod, is also rather simple and discussed in \cite{B&Hb,Pomeau}. The even eigenfunctions are of the form $u(x)=\cos (\lambda x) + \sin \lambda \cosh (\lambda x)/\sinh \lambda$, where $\lambda$ is any solution of the transcendental equation $\tan \lambda + \tanh\lambda=0$. The odd eigenfunctions are of the form $u(x)=\sin (\lambda x) - \sin \lambda \sinh (\lambda x)/\sinh \lambda$, where $\lambda$ is any solution of $\tan \lambda - \tanh\lambda=0$. The allowed values of $\lambda$ cannot be written in closed form, although they are easily studied numerically and asymptotically. For $N=3$, a special symmetry simplifies the problem \cite{B&Hb,Zoia}. As a result, the eigenvalues of the whole (odd and even) spectrum are exactly $\lambda_{2j}=(j\pi)^6$ and $\lambda_{2j+1}=((j+1/2)\pi+\delta_j)^6$ with $\delta_j \sim 4(-1)^{j +1}e^{-(\pi \sqrt{3}/2)(2j+1)}$ for large $j$ \cite{B&Hb}. Finally, for $N>3$ it was shown that the eigenvalues of the whole spectrum are asymptotically known \cite{B&Hb}. More precisely, there exists a $j_0$ such that for $j>j_0$
\begin{equation}
\fl \lambda _j  = \left\{ {\frac{\pi}{2} \left( {j + {\textstyle{1
\over 2}}} \right) + \frac{{\left( { - 1} \right)^{j + 1} }}{{\sin
^2 {\textstyle{\pi  \over {2N}}}}}e^{ - \pi \left( {j +
{\textstyle{1 \over 2}}} \right)\sin {\textstyle{\pi  \over N}}}
\cos \left[ {\pi \left( {j + {\textstyle{1 \over 2}}} \right)\cos
{\textstyle{\pi \over N}}} \right] + O\left( {e^{ - 2\pi j\sin
{\textstyle{\pi \over N}}} } \right)} \right\}^{2N}
 \label{eq:BHeven} \,,
\end{equation}
for $N$ even, while for odd $N$'s a slightly different expression exists
\begin{equation}
\fl \lambda _j  = \left\{ {\frac{\pi}{2} j + \frac{{\left( { - 1}
\right)^{j} }}{{\sin ^2 {\textstyle{\pi  \over {2N}}}}}e^{ - \pi
j\sin {\textstyle{\pi  \over N}}} \sin \left[ {\pi j\cos
{\textstyle{\pi  \over N}}} \right] + O\left( {e^{ - \pi j\sin
{\textstyle{{2\pi } \over N}}} } \right)} \right\}^{2N}
 \label{eq:BHodd} \,.
\end{equation}
However $j_0$ can be in practice rather large and thus the lower part of the spectrum, which is often the most interesting in physical problems, cannot be explored using (\ref{eq:BHeven},\ref{eq:BHodd}). Note also that Eqs.~(\ref{eq:BHeven},\ref{eq:BHodd}) do not provide any information about the eigenfunctions.

Recently, it was shown \cite{B&Ha,B&Hb,Pomeau} that for large $N$ the eigenfunction with the lowest eigenvalue (so called ``the ground state" borrowing the terminology of quantum mechanics) is given by $v_0(x) = \left(1 - x^2\right)^N $ and its corresponding eigenvalue is $\lambda_0 \sim \sqrt 2 \left( {2N} \right)!$ with corrections which are of order $1/N$ compared to the leading order term. Actually, in \cite{Pomeau} many more terms were calculated for the lowest eigenvalue, namely
\begin{equation}
\fl \lambda_0  = \sqrt 2 \left( {2N} \right)!\left[ {1 -
\frac{3}{{16N}} + \frac{{25}}{{512N^2 }} + \frac{{375}}{{8192N^3 }}
- \frac{{8197}}{{524288N^4 }} + O\left( {\frac{1}{{N^5 }}} \right)}
\right]
 \label{eq:lminP} \,.
\end{equation}
Also, it was shown in \cite{Pomeau} that the first excited eigenvalue is
\begin{equation}
\lambda_1 = \sqrt 2 (2N)! 8 N^2 \left[1- \frac{15}{16N} +O\left(\frac{1}{N^2}\right)\right]\,.
\end{equation}

The present situation calls for more results concerning the spectrum of $\Delta^N$. In this paper, we show that in the large $N$ limit, the eigenfunction of $\Delta^N$ with absorbing BC can be written as certain associated Legendre polynomials. This also suggests, and is shown to be true below, that this basis is a good choice for diagonalizing the differential operator $\Delta^N$ also when $N$ is not large. To be more precise, if we consider the normalized associated Legendre polynomials
\begin{equation}
\hat P_{2N + k}^{2N} \left( x \right) \equiv \sqrt {\frac{{4N + 2k +
1}}{2}\frac{(k)!}{(4N + k)!}} P_{2N + k}^{2N} \left( x \right)
 \label{eq:ALP1} \, ,
\end{equation}
we show that the eigenfunctions of even spectrum are just given by
\begin{equation}
v_j (x) = \hat P_{2N + 2j}^{2N} \left( x \right) + O\left(
{\frac{1}{N}} \right) \quad \quad {\rm{for}} \quad j = 0,1,2, \ldots
 \label{eq:spec1} \, .
\end{equation}
This result has a simple origin. In order to get some insight into it, recall that the associated Legendre polynomials form an orthogonal basis for functions in the interval $[-1,1]$, that is
\begin{equation}
\int_{-1}^1 {\hat P_{2N + i}^{2N}(x) \hat P_{2N + j}^{2N}(x) dx}
=\delta_{ij} \quad \quad \quad \quad i,j = 0,1,2, \ldots
 \label{eq:orhonormality} \, ,
\end{equation}
and they all obey the BC (\ref{eq:BC}). This means that without loss of generality any eigenfunction of the aforementioned problem can be written as a linear combination of them. In particular, any normalized even function can be expanded using only even eigenfunction as follows (similarly, for the odd spectrum one needs the odd eigenfunction)
\begin{equation}
U(x) = \sum\limits_{j = 0}^\infty  {U^{(j)} \hat P_{2N + 2j}^{2N}
\left( x \right)}
 \label{eq:evenALP} \, ,
\end{equation}
with $\left\| {U(x)} \right\|^2  = \sum\limits_{j = 0}^\infty{\left[ {U^{(j)} } \right]^2 }= 1$. It is also well known \cite{Courant,Pomeau} that the ground state can be obtained from a variational approach, namely it is obtained as the minimum of the functional
\begin{equation}
\lambda\{U\} \equiv \frac{{\int\limits_{ - 1}^1 {U\left( x
\right)\left[ {\Delta ^N U\left( x \right)} \right]dx}
}}{{\int\limits_{ - 1}^1 {\left[ {U\left( x \right)} \right]^2 dx}}}
 \label{eq:variational} \,
\end{equation}
over all functions $U(x)$ that obey the BC (\ref{eq:BC}). In \cite{B&Ha,Pomeau}, it was proved  that the ground state in the large $N$ limit is simply given by $v_0(x)=\left(1 - x^2\right)^N$, which is just $\hat P_{2N}^{2N}(x)$ up to a numerical constant. This result is already consistent with Eq.~(\ref{eq:spec1}) for $j=0$. The next step is to look for the first excited state, which can also be obtained from the variational approach, as it is just the minimum of $\lambda\{U\}$ as given by Eq.~(\ref{eq:variational}) over all functions $U(x)$ that obey the BC (\ref{eq:BC}) and are orthogonal to $v_0(x)$. This means that in full generality, $v_1(x)$ can be expanded in the form
\begin{equation}
v_1(x) = \sum\limits_{j = 1}^\infty  {v_1^{\left( j \right)} \hat
P_{2N + 2j}^{2N} \left( x \right)}  \quad \quad  {\rm with} \quad
\quad \sum\limits_{j = 1}^\infty  {\left[ v_1^{(j)} \right]^2 }= 1
 \label{eq:v1} \, .
\end{equation}
in the large $N$ limit, since the set $\left\{ {\hat P_{2N + 2j}^{2N} \left( x \right)} \right\}_{j = 1 \ldots \infty }$ spans the (even) subspace which is orthogonal to $v_0(x)$.

Before providing a full proof below, we present a heuristic argument which can give some insight into the result (\ref{eq:spec1}). We claim that every zero crossing of a function $U(x)$ that obeys the BC (\ref{eq:BC}) adds a factor of order $N$ to the functional $\lambda\{U\}$ (\ref{eq:variational}) due to the presence of the large derivative, for large $N$. Recall that $\hat P_{2N + k}^{2N}$ has exactly $k$ roots in the interval $[-1,1]$ \cite{G&R}. This means that in the general expansion (\ref{eq:v1}), the best candidate to provide a minimum for $\lambda\{U\}$ is simply $v_1^{\left( 1 \right)}  = 1$ and $v_1^{\left( {j \geq 2} \right)} = O(1/N)$, as it has the minimal number of roots in the interval. Finally, this heuristic argument that yileds $v_1(x)$ can be recursively generalized to any $v_j(x)$, and the result (\ref{eq:spec1}) is obtained.

In the rest of the paper, we first provide a proof of the statement (\ref{eq:spec1}). We use this result to obtain some explicit expressions for the eigenfunctions and eigenvalues in the large $N$ limit. Then, we will show that expressing $\Delta^N$ in the basis of the associated Legendre polynomials does not only diagonalize it for $N\rightarrow \infty$, but is also a very useful basis for numerically evaluating its spectrum for any $N$. Finally, we discuss possible extensions and implications of this work.

\section{The Spectrum in the large $N$ limit}

In order to prove the statement (\ref{eq:spec1}), we start by expressing the matrix elements $\hat \Delta _{m,j}^N$ of the operator $\Delta^N$ in the basis of the associated Legendre polynomials~(\ref{eq:ALP1})
\begin{equation}
 \hat \Delta _{m,j}^N  \equiv \int\limits_{ - 1}^1 {\hat P_{2N + 2m}^{2N} \left( x \right)\left[ {\Delta ^N \hat P_{2N + 2j}^{2N} \left( x \right)} \right]dx}
\label{eq:matrix} \, ,
\end{equation}
After some algebraic manipulations which are summarized in Appendix A, one finds
\begin{eqnarray}
\fl
\begin{array}{l}
 \hat \Delta _{m,j}^N  = \sqrt {{\textstyle{{\frac{\left( {2j} \right)!{\left( {4N + 2m} \right)!}}{{\left( {2m} \right)!\left( {4N + 2j} \right)!}}}}}} {\textstyle{{\sqrt {\left( {4N + 4m + 1} \right)\left( {4N + 4j + 1} \right)} \left( { - 1} \right)^{N + j} } \over {2^{2N + 1} \Gamma \left( {N + {\textstyle{1 \over 2}}} \right)}}}\sum\limits_{i = 0}^j {{\textstyle{{\left( { - 4} \right)^i \left( {2N + 2i} \right)!\Gamma \left( {2N + j + i + {\textstyle{1 \over 2}}} \right)\Gamma \left( {i + {\textstyle{1 \over 2}}} \right)} \over {\left[ {\left( {2i} \right)!} \right]^2 \left( {j - i} \right)!\Gamma \left( {N + i + {\textstyle{3 \over 2}}} \right)}}}}  \\
  \quad\quad \times {}_3F_2 \left( {\begin{array}{*{20}c}
   {i - j, - N,2N + i + j + {\textstyle{1 \over 2}}}  \\
   {i + {\textstyle{1 \over 2}},i + 1}  \\
\end{array};1} \right){}_3F_2 \left( {\begin{array}{*{20}c}
   {2N + m + {\textstyle{1 \over 2}}, - m,N + 1}  \\
   {2N + 1,N + i + {\textstyle{3 \over 2}}}  \\
\end{array};1} \right) \\
 \end{array}
 \label{eq:matrix1} \, ,
\end{eqnarray}
where $\Gamma(u)$ is Euler's Gamma function, and ${}_3F_2 \left( {\begin{array}{*{20}c} {a,b,c}\\{d,e}\\ \end{array};x} \right)$ is the generalized hypergeometric function \cite{G&R}. Note that the matrix $\hat \Delta^N $ is symmetric ($\hat \Delta _{m,j}^N  = \hat \Delta _{j,m}^N$), as can be shown by
integrating by parts (\ref{eq:matrix}) $N$ times and keeping in mind that the associated Legendre polynomials obey the BC (\ref{eq:BC}).

In the following, we will need two results regarding this infinite dimensional matrix. First, we show in Appendix B that the matrix elements behave asymptotically for large $N$ as
\begin{equation}
\hat \Delta _{m,j}^N  = \left( { - 1} \right)^N \sqrt 2 \left( {2N}
\right)!\left( {4N} \right)^{2j} \sqrt {\frac{{\left( {2m}
\right)!}}{{\left( {2j} \right)!}}} \frac{{2^{j - m} }}{{\left( {m -
j} \right)!\left( {2j} \right)!}}\left[ {1 + O\left( {\frac{1}{N}}
\right)} \right]
 \label{eq:deltajm} \, ,
\end{equation}
for $m \ge j$. Retaining one more term for the diagonal elements gives
\begin{equation}
\hat \Delta _{j,j}^N  = \left( { - 1} \right)^N \sqrt 2 \left( {2N}
\right)! \frac{{\left(4N\right)^{2j}
}}{{\left( {2j} \right)!}}\left[ {1 -
\frac{{3 + 4j + 8j^2 }}{{16N}} + O\left( {\frac{1}{{N^2 }}} \right)}
\right]
 \label{eq:deltajj} \, .
\end{equation}
This means that to leading order the matrix has the following structure
\begin{equation}
\fl \hat \Delta ^N \sim \left( { - 1} \right)^N \sqrt 2 \left( {2N}
\right)!\left( {\begin{array}{*{20}c}
   1 & {{\textstyle{1 \over {\sqrt 2 }}}} & {\sqrt {{\textstyle{3 \over 8}}} } &  \cdots  & {{\textstyle{{\sqrt {\left( {2j} \right)!} } \over {2^j j!}}}} &  \cdots   \\
   {{\textstyle{1 \over {\sqrt 2 }}}} & {8N^2 } & {8\sqrt 3 N^2 } &  \cdots  & {{\textstyle{{\sqrt {\left( {2j} \right)!} } \over {2^{j - 7/2} \left( {j - 1} \right)!}}}N^2 } &  \cdots   \\
   {\sqrt {{\textstyle{3 \over 8}}} } & {8\sqrt 3 N^2 } & {{\textstyle{{32} \over 3}}N^4 } &  \cdots  & {{\textstyle{{\sqrt {\left( {2j} \right)!} } \over {2^{j - 11/2} 3^{{3 \mathord{\left/
 {\vphantom {3 2}} \right.
 \kern-\nulldelimiterspace} 2}} \left( {j - 2} \right)!}}}N^4 } &  \cdots   \\
    \vdots  &  \vdots  &  \vdots  &  \ddots  &  \vdots  &  \cdots   \\
   {{\textstyle{{\sqrt {\left( {2j} \right)!} } \over {2^j j!}}}} & {{\textstyle{{\sqrt {\left( {2j} \right)!} } \over {2^{j - 7/2} \left( {j - 1} \right)!}}}N^2 } & {{\textstyle{{\sqrt {\left( {2j} \right)!} } \over {2^{j - 11/2} 3^{{3 \mathord{\left/
 {\vphantom {3 2}} \right.
 \kern-\nulldelimiterspace} 2}} \left( {j - 2} \right)!}}}N^4 } &  \cdots  & {{\textstyle{{2^{4j} } \over {\left( {2j} \right)!}}}N^{2j} } &  \cdots   \\
    \vdots  &  \vdots  &  \vdots  &  \vdots  & {} &  \ddots   \\
\end{array}} \right)
 \label{eq:matrixrep} \, .
\end{equation}

\subsection{Proof for the diagonalization}

We will now prove by induction that in the large $N$ limit the spectrum of this matrix is just its diagonal elements so that it is essentially diagonal in this basis to order $1/N$ and thanks to some extra structure, this is true actually up to order $1/N^2$.

Let us suppose that this statement is true for the submatrix of size $K$ defined as $\hat \Delta _{m,j}^{N,K}  \equiv \left\{ {\hat \Delta _{m,j}^N } \right\}_{m,j = 0, \cdots ,K}$ and then we will show that it holds for
$\hat \Delta _{m,j}^{N,K+1}$ as well. Take the trace of the finite $\left( {K+ 1} \right) \times \left( {K + 1} \right)$ submatrix. Using Eq.~(\ref{eq:deltajj}) it is given explicitly by
\begin{equation}
\fl Tr\hat \Delta ^{N,K + 1}  = \left( { - 1} \right)^N \sqrt 2
\left( {2N} \right)!\sum\limits_{j = 0}^{K + 1} {
\frac{{(4N)^{2j} }}{{\left( {2j} \right)!}}}  = \left( { - 1} \right)^N
\sqrt 2 \left( {2N} \right)! \frac{{\left(4N\right)^{2\left( {K + 1}
\right)} }}{{\left( {2K + 2} \right)!}}\left[ {1 + O\left(
{\frac{1}{{N^2 }}} \right)} \right]
 \label{eq:trace} \, .
\end{equation}
This simply means that $\hat \Delta _{K + 1,K + 1}^{N,K+1}  - Tr\hat \Delta ^{N,K + 1}  = O\left( {\frac{1}{{N^2 }}} \right)$. Note also that $\hat \Delta _{K + 1,K + 1}^{N}$ is the only term in the submatrix $\hat \Delta ^{N,K+1}$ that is of order $\left( {2N} \right)!N^{2K + 2}$. This together with the fact that the trace is just the sum of all the eigenvalues, means that $\hat \Delta _{K + 1,K + 1}^N$ is an eigenvalue of $\hat \Delta ^{N,K+1}$, and it is the maximal one. It is easy to check that the column vector $v_{K + 1}  \equiv \left( {0, \cdots ,0,1} \right)^T$ is the eigenvector that corresponds to this eigenvalue, up to corrections of order $O\left( {1/N^2 } \right)$, since
\begin{eqnarray}
\fl \hat \Delta ^{N,K + 1} v_{K + 1} &=& \left( { - 1} \right)^N \sqrt 2 \left( {2N} \right)!\left( {{\textstyle{{\sqrt {\left( {2K + 2} \right)!} } \over {2^{K + 1} \left( {K + 1} \right)!}}}, \cdots ,{\textstyle{{\sqrt {\left( {2K + 2} \right)\left( {2K + 1} \right)} \left( {4N} \right)^{2K} } \over {2\left( {2K} \right)!}}},{\textstyle{{\left( {4N} \right)^{2\left( {K + 1} \right)} } \over {\left( {2K + 2} \right)!}}}} \right)^T   \nonumber \\
 &=& \left( { - 1} \right)^N \sqrt 2 \left( {2N} \right)!{\textstyle{{\left( {4N} \right)^{2\left( {K+ 1} \right)} } \over {\left( {2K + 2} \right)!}}}\left( {0, \cdots ,0,1} \right)^T \left[ {1 + O\left( \frac{1}{N^2 } \right)} \right]
 \label{eq:eigenvector} \, .
\end{eqnarray}
We have succeeded in finding one eigenvalue and one eigenvector, and in order to complete the diagonalization we need to consider the upper left $K \times K$ submatrix. However, this is nothing but $\hat \Delta ^{N,K}$ which is by assumptions diagonal in this basis. QED.

\subsection{Higher order results}

Note that all the arguments of the proof above are valid up to corrections of order $O\left( {1/N^2 } \right)$. This implies that the diagonal elements of $\hat \Delta^N$ are indeed the eigenvalues up to $O\left( {1/N^2 } \right)$. This fact becomes useful when considering higher order corrections, since we know that Eq.~(\ref{eq:deltajj}) already gives the first subleading corrections, without any need to diagonalize the matrix.

The same reasoning and a quick look at Eq.~(\ref{eq:eigenvector}) lead to the conclusion that in order to get the first correction for the eigenfunction (which is of order $1/N^2$) we need to diagonalize blocks of $3 \times 3$ (with the exception of the ground state where the $2 \times 2$ upper-left submatrix suffices). Thanks to the special structure of the asymptotic behavior of the matrix, it is sufficient to know the ratio $\hat\Delta^N_{j,j+1}/\hat\Delta^N_{j+1,j+1}$ up to order $1/N$, namely
\begin{equation}
\frac{\hat\Delta^N_{j,j+1}}{\hat\Delta^N_{j+1,j+1}} =  - \frac{{\left[ {\left(
{2j + 2} \right)\left( {2j + 1} \right)} \right]^{{3 \mathord{\left/
{\vphantom {3 2}} \right. \kern-\nulldelimiterspace} 2}}
}}{{32}}\left( {1 - \frac{{3 + 4j}}{{8N}}} \right) + O\left(
{\frac{1}{{N^2 }}} \right)
 \label{eq:ratio} \, ,
\end{equation}
and so we get
\begin{eqnarray}
\fl v_j \left( x \right) &=&  \hat P_{2N + 2j}^{2N} \left( x \right)\nonumber\\
\fl &+&\frac{1}{32N^2}\left\{{\textstyle{{\left[ {\left( {2j}
\right)\left( {2j - 1} \right)} \right]^{{3 \mathord{\left/
 {\vphantom {3 2}} \right.
 \kern-\nulldelimiterspace} 2}} }}}\hat P_{2N + 2j - 2}^{2N} \left( x \right)  - {\textstyle{{\left[ {\left( {2j + 2} \right)\left( {2j + 1} \right)} \right]^{{3 \mathord{\left/
 {\vphantom {3 2}} \right.
 \kern-\nulldelimiterspace} 2}} }}}\hat P_{2N + 2j + 2}^{2N} \left( x \right)\right\} + O\left( {\frac{1}{{N^3 }}} \right)
 \label{eq:eigenvector-corr} \, .
\end{eqnarray}

In turn this allows to know the next two subleading orders for the eigenvalues, that is up to order $1/N^3$. After some algebra we get
\begin{eqnarray}
 \fl \lambda _j  &=& \sqrt 2 (4N)^{2j} \frac{{ \left( {2N} \right)! }}{{\left( {2j} \right)!}}\left[ {1 - \frac{{3 + 4j + 8j^2 }}{{16N}} + \frac{{75 + 344j + 672j^2  + 832j^3  + 192j^4 }}{{1536N^2 }} + } \right. \nonumber \\
 \fl &&\left. { + \frac{{1125 + 5028j + 11960j^2  + 5120j^3  + 4416j^4  - 5888j^5  - 512j^6 }}{{24576N^3 }} + O\left( {\frac{1}{{N^4 }}} \right)} \right]
 \label{eq:lambdaN} \, .
\end{eqnarray}
Repeating the same method, we can also obtain even higher order terms for specific states without serious difficulties. For example, the first two eigenvalues are given by
\begin{eqnarray}
 \fl \lambda _0  = \sqrt 2 \left( {2N} \right)!\left[ {1 - \frac{3}{{16N}} + \frac{{25}}{{512N^2 }} + \frac{{375}}{{8192N^3 }} - \frac{{8197}}{{524288N^4 }} - \frac{{694941}}{{8388608N^5 }} + O\left( {\frac{1}{{N^6 }}} \right)} \right] \\
 \fl \lambda _1  = 8 \sqrt 2 N^2\left( {2N} \right)!\left[ {1 - \frac{{15}}{{16N}} + \frac{{705}}{{512N^2 }} + \frac{{7083}}{{8192N^3 }} - \frac{{{\rm{963989}}}}{{{\rm{524288}}N^4 }} - \frac{{{\rm{68014113}}}}{{{\rm{8388608}}N^5 }} + O\left( {\frac{1}{{N^6 }}} \right)} \right]
 \label{eq:lambda01} \, ,
\end{eqnarray}
which are consistent with the results of \cite{Pomeau}, where these quantities were calculated up to $1/N^4$ for the ground state, and up to $1/N$ for the first excited state.

Recall that these results are exact when $N\to \infty $. However, for a finite $N$, there always exists a certain $j$ for which these first terms are not sufficient, and in order to obtain the eigenvalues/eigenvectors one needs to know more terms of the expansion. This should have been expected since as has been conjectured in \cite{Pomeau} and proved in \cite{B&Hb} there always exists a $j_0$ (which can become very large actually) such that for $j > j_0$ the eigenvalues approach asymptotically $\left( {\frac{\pi } {2} + \pi j} \right)^{2N}$ for odd $N$ and $\left( {\frac{3\pi } {4} + \pi j} \right)^{2N}$ for even $N$. Such a transition to``simple" eigenvalues necessarily implies that many terms in the large $N$ series should be retained as we get to higher excited states.

So far, the results mentioned above apply only to the even sector of the spectrum. It turns out that the odd part can be obtained by simply transforming $j \rightarrow j+\frac{1}{2}$ in all the formulas for the even part, namely Eqs.~(\ref{eq:spec1}) and (\ref{eq:deltajm})-(\ref{eq:lambdaN}). As a consequence the proof
for the even part can be simply repeated for the odd part. Therefore, we essentially fully cover the two sectors of the spectrum for large $N$. An interesting conclusion, that was conjectured already in the introduction and is now proven, is that in the spectrum each extra zero-crossing of the eigenfunction ``costs" a factor $N$ in its corresponding eigenvalue in the large-$N$ limit.

\section{Numerical approach}

The proof presented in the previous section shows that the orthonormal set $\left\{ {\hat P_{2N + 2j}^{2N} \left( x \right)} \right\}$ diagonalizes the operator $\Delta^N$ for large $N$. However, from a more general point of view, this orthonormal set is a complete basis for even functions that obey the BC (\ref{eq:BC}) and
therefore, by representing $\Delta^N$ in this basis we loose no information. For that reason, a direct diagonalization of the matrix $\hat \Delta_{m,j}^N$ should provide the eigensystem of $\Delta^N$
for any $N$, even though it is not clear {\em a priori} if this approach is efficient. On the other hand, since we proved that this basis diagonalizes $\Delta^N$ in the large $N$ limit, we expect it to be a good starting point for diagonalization for any $N$. Indeed, we now show that one can get very accurate results for finite  $N$ by numerically treating the matrix $\hat\Delta_{m,j}^N$ (\ref{eq:matrix}).

\begin{figure}[ht]
\begin{center}
\includegraphics[width=7cm]{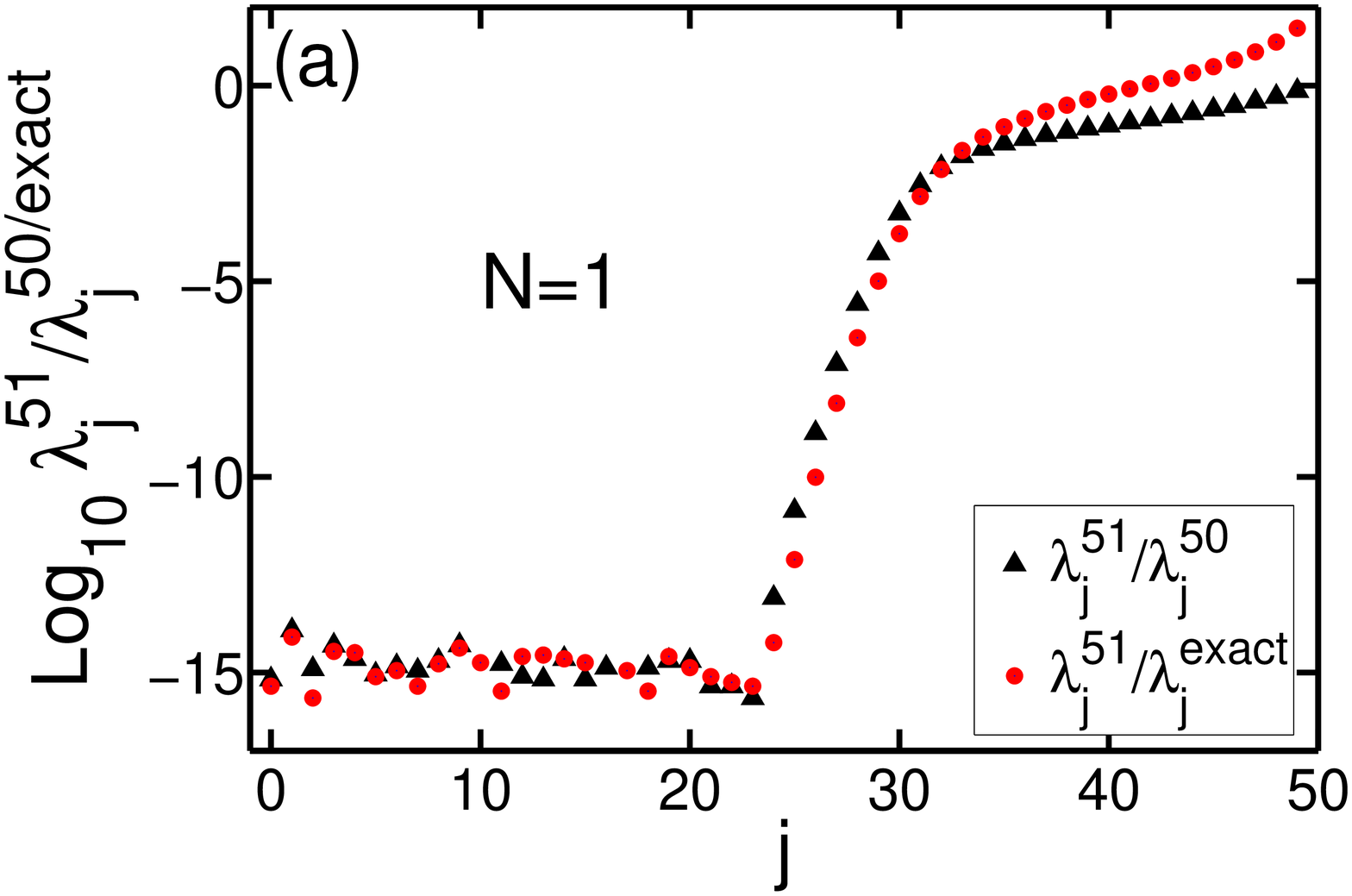}\includegraphics[width=7cm]{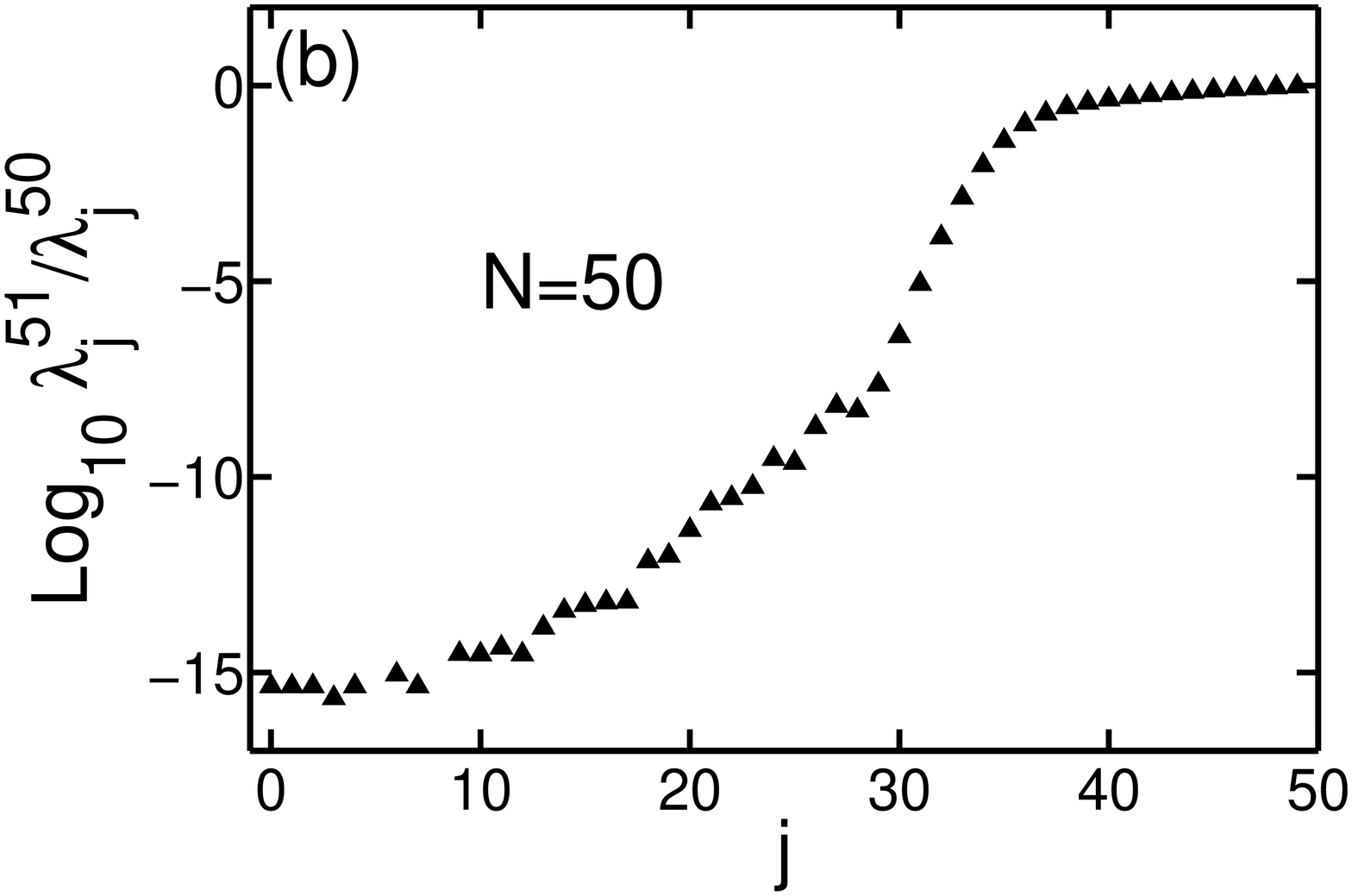}
\caption{A plot of the estimated accuracy for the first $50$ eigenvalues obtained from a truncated $51 \times 51$ matrix, i.e. $\hat \Delta^{N,51}_{j,m}$, for (a) $N=1$ and (b) $N=50$. In both cases we assess the accuracy by comparing to the first $50$ eigenvalues of $\hat \Delta^{N,50}_{j,m}$. For $N=1$ we also compare to the exact result $\lambda_j=\left[\pi\left(j+\frac{1}{2}\right)\right]^2$. Note that the two estimations agree quite well with each other.}
 \label{fig:accuracy}
\end{center}
\end{figure}

In order to show the high rate of convergence we can consider the case $N = 1$ directly. Note that for $N=1$ we know exactly the whole spectrum, which is $\left\{ {\cos \left( {{\textstyle{\pi
\over 2}} + n\pi } \right)} \right\}_{n = 0 \ldots \infty }$ with its corresponding eigenvalues $\left( {{\textstyle{\pi  \over 2}} + n\pi } \right)^2 $. The following expansion holds
\begin{equation}
\cos \left( {{\textstyle{\pi  \over 2}} + n\pi } \right) =
\sum\limits_{i = 0}^\infty  {c_i P_{2 + 2i}^2(x)}
 \label{eq:cos} \, ,
\end{equation}
with
\begin{eqnarray}
 c_i  &=& \frac{{(-1)^i\left( {2i} \right)!}}{{\left( {2i + 4} \right)!}} (4i + 5)\left\{ {2{{{\left( {2n + 1} \right)\pi }}}j'_{2i + 2} \left( {{\textstyle{{\left( {2n + 1} \right)\pi } \over 2}}} \right) + } \right. \nonumber \\
 && \left. {+ \left[ {2 + \pi ^2 \left( {n + {\textstyle{1 \over 2}}} \right)^2 } \right]j_{2i + 2} \left( {{\textstyle{{\left( {2n + 1} \right)\pi } \over 2}}} \right) + \pi ^2 \left( {n + {\textstyle{1 \over 2}}} \right)^2 j''_{2i + 2} \left( {{\textstyle{{\left( {2n + 1} \right)\pi } \over 2}}} \right)} \right\} \nonumber  \\
 &&  \sim \frac{{\left( { - 1} \right)^i }}{i}\frac{{e^{2i + {\textstyle{5 \over 2}}} \left[ {\left( {n + {\textstyle{1 \over 2}}} \right)\pi } \right]^{2i + 2} }}{{2^{2i + {\textstyle{7 \over 2}}} \left( {2i + {\textstyle{5 \over 2}}} \right)^{2i + 3} }}
 \label{eq:ck} \, .
\end{eqnarray}
The coefficients $c_i$ decay faster than exponentially to zero for large $i$, which means that one practically needs a small number of coefficients in order to get a good approximation of the real eigenfunctions. Note that when diagonalizing the matrix $\hat \Delta^1_{j,m}$ ((Eq.~(\ref{eq:matrix}) with $N=1$), we exactly calculate the coefficients $c_i$, and therefore it is enough to consider just a finite submatrix approximation to the full matrix $\hat \Delta_{j,m}$  in order to get a good result for the eigensystem. A simple thumb rule of taking a finite submatrix, diagonalizing it and then keeping only the first half eigenvalues yields result with good accuracy. In Fig.~\ref{fig:accuracy} we present two examples where we evaluate the accuracy of this approach for the eigenvalues for $N=1$ and $N=50$. In both cases we assess the accuracy by comparing the eigenvalues obtained from a truncated $51 \times 51$ matrix, i.e. $\hat \Delta^{N,51}_{j,m}$, to those obtained from $\hat \Delta^{N,50}_{j,m}$. For $N=1$, since we know the exact eigenvalues, namely $\lambda_j=\left[\pi\left(j+\frac{1}{2}\right)\right]^2$, we can
also have an absolute check on the accuracy. As one can see the simple thumb rule of keeping only the first half eigenvalues amounts to a relative error of at least $\sim 10^{-10}$. Keeping in mind that the eigenvalues grow very fast with $N$ and $j$ (here we only look at ratios in order to have a better evaluation) this is reassuring.

\begin{figure}[ht]
\centerline{\includegraphics[width=7cm]{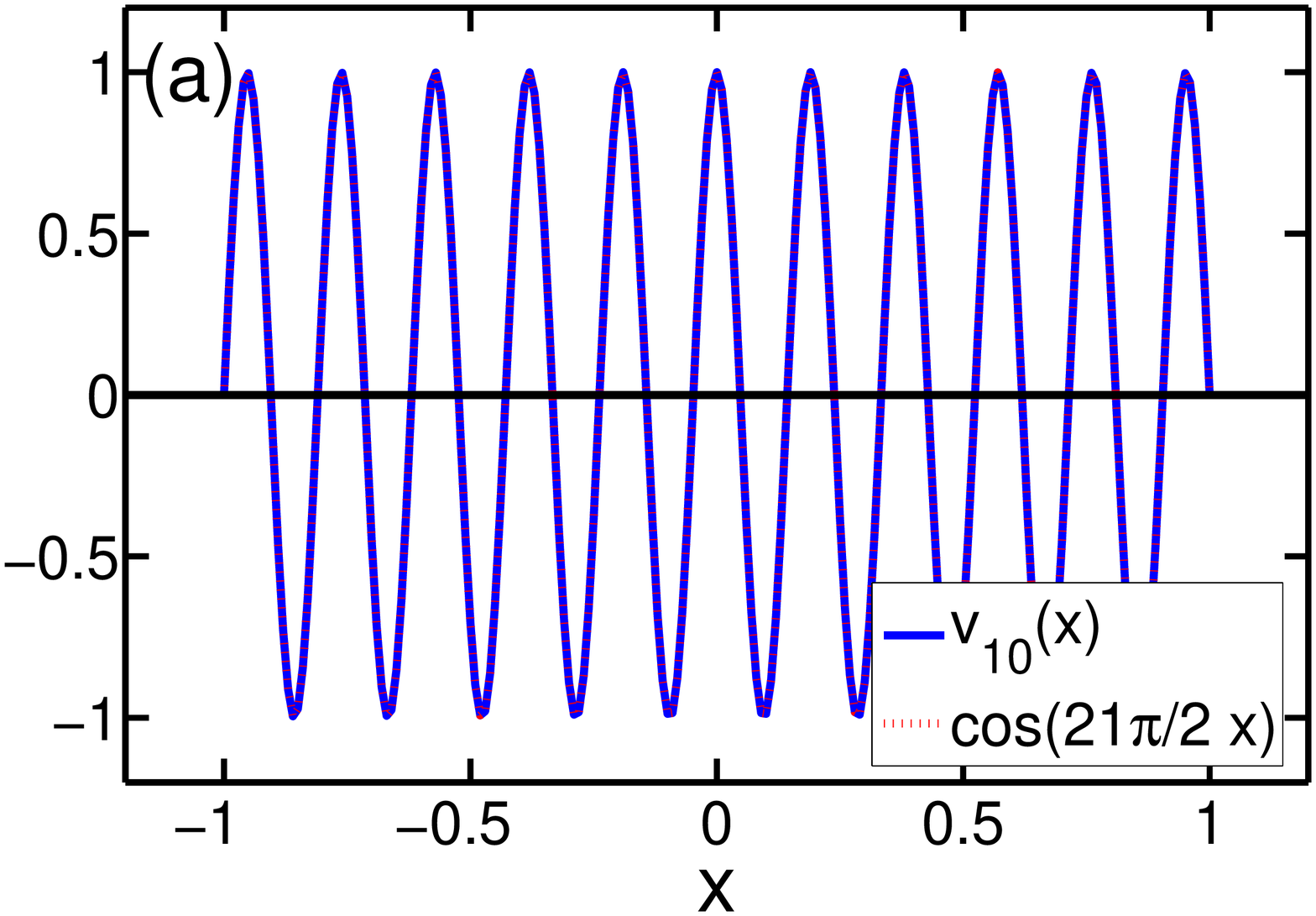}\includegraphics[width=7cm]{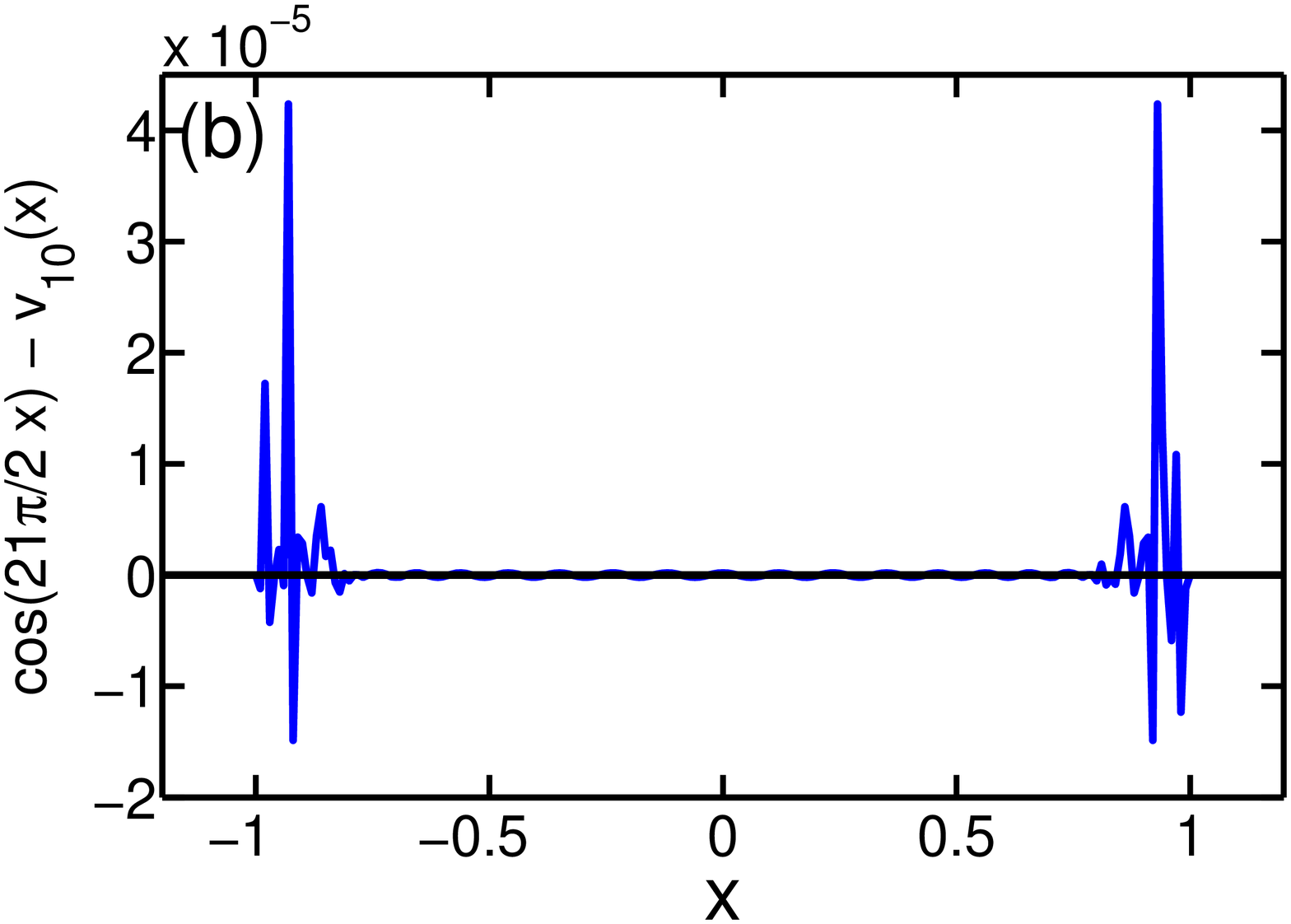}}
\caption{A comparison between the $10^{th}$ (even) excited state obtained from the numerical scheme using a truncated matrix of $50 \times 50$, i.e. $\hat \Delta^{N,50}_{j,m}$, denoted here $v_{10}(x)$ and the exact eigenfunction $\cos(\frac{21 \pi}{2}x)$ for the case $N=1$. In (a) we plot the two functions, where they seem to agree very well, while in (b) we plot the difference of the two in order to allow for a more quantitative evaluation. Most of the error is concentrated near the boundaries, where it is of the order of $10^{-5}$. The integral of the absolute value of the difference is $\int_{-1}^1{\left|v_{10}(x)-\cos(\frac{21 \pi}{2}x)\right| dx}=3.43 \times 10^{-6}$.}
 \label{fig:10thfunction}
\end{figure}

Concerning the eigenfunctions, the situation is of course more delicate. In order to get sufficient convergence throughout the interval $[-1,1]$, we had to keep around $1/3$ of the states. In Fig. \ref{fig:10thfunction}, we compare the $10^{th}$ excited state $v_{10}(x)$ with the exact result $\cos(\frac{21 \pi}{2}x)$ for $N=1$. As can be seen, the two functions agree very well and the difference reveals a discrepancy of order $10^{-5}$, which is mainly concentrated close to the boundaries.

The latter discussion has an important implication. If the above procedure converges rapidly for $N=1$ then it certainly does so for higher values of $N$, which means that this numerical
procedure can be useful in general. Having established the fact that the numerical scheme yields results with controllable accuracy, we can use the outcome in order to assess the various analytical
results obtained so far. In Table \ref{tab:comaprison}, we compare the numerical results for the eigenvalues obtained from the truncated matrix of $50 \times 50$ (i.e., $\hat \Delta^{N,50}_{j,m}$) to the large $N$ $4^{th}$ order formula we obtained in Eq.~(\ref{eq:lambdaN}) and to the asymptotic result $\left[\pi(j+3/4)\right]^{2N}$ of \cite{B&Hb} for $N=50$. First, notice how fast the eigenvalues grow already for $N=50$. Second, one can see the large $N$ formula (\ref{eq:lambdaN}) describes well the low states and starts deviating around $j \sim 12$ where they become negative. Table~\ref{tab:comaprison} also shows that when $N$ becomes large, the critical $j_0$ above which the asymptotic result of \cite{B&Hb} is valid becomes large itself.
\begin{table}
\begin{indented}
\lineup
\item[]\begin{tabular}{@{}*{4}{l}}
\br $j$&$\lambda^{num}_j/\sqrt{2}(2
N)!$&$\lambda^{num}_j/\lambda^{4th}_j$&\m$\lambda^{num}_j/\left[\pi(j+3/4)\right]^{2N}$\cr
\mr
 $1$  & $0.99627$               & $1.$          & $7.90282\times 10^{120}$ \cr
 $3$  & $6.33108\times 10^7$    & $0.999996$    & $1.87819\times 10^{72}$ \cr
 $5$  & $5.35907\times 10^{13}$ & $1.00007$     & $2.91917\times 10^{54}$ \cr
 $7$  & $6.04205\times 10^{18}$ & $1.00309$     & $1.80442\times 10^{44}$ \cr
 $9$  & $1.7643\times 10^{23}$  & $1.0346$      & $2.82683\times 10^{37}$ \cr
 $11$ & $1.86415\times 10^{27}$ & $1.27419$     & $3.42899\times 10^{32}$ \cr
 $13$ & $8.751\times 10^{30}$   & $-42.7013$    & $6.25995\times 10^{28}$ \cr
 $15$ & $2.10073\times 10^{34}$ & $-0.337001$   & $7.05830\times 10^{25}$ \cr
 $20$ & $5.86444\times 10^{41}$ & $-0.0158765$  & $4.14045\times 10^{20}$ \cr
 $25$ & $1.31862\times 10^{48}$ & $-0.00182839$ & $1.47275\times 10^{17}$ \cr \br
\end{tabular}
\end{indented}
\caption{\label{tab:N=50} A comparison between the different results for the eigenvalues of the even spectrum for $N=50$. $\lambda^{num}_j$ that results from the numerical procedure by using a truncated matrix of $50 \times 50$, $\lambda^{4th}$ that come from Eq.~(\ref{eq:lambdaN}) and the asymptotic result $\left[\pi(j+3/4)\right]^{2N}$ \cite{B&Hb}. As can be seen, the first $11$-$12$ eigenvalues are reasonably captured by $\lambda^{4th}$, while the asymptotic formula is still far from giving a good estimate even for $j=25$. The negative values reflect the fact that $\lambda_j^{4th}$ is no longer a good approximation for $j>12$.}
\label{tab:comaprison}
\end{table}

\section{Summary and discussion}

In this paper we calculate the spectrum of the $N^{th}$ power of the Laplace operator in the large $N$ limit. We obtain explicit expressions for all the eigenfunctions up to order $1/N^2$ and for all the eigenvalues up to order $1/N^3$, for both the even and the odd parts of the spectrum. In addition we show the general structure of the series which provides corrections to the eigenvectors of order $1/N^{2m}$ by diagonalizing blocks of $(2m+1) \times (2m+1)$ for large $N$, and keeping consistently all term up order $1/N^{2m}$ in the matrix elements that contribute. The expansion we obtain is consistent with some previous known results for the ground state \cite{B&Ha,Pomeau} and for the first excited state \cite{Pomeau}. In addition, we show that this approach can be useful for small $N$'s, where a direct diagonalization of the matrix representation of the operator $\Delta^N$ in the basis of the associated Legendre polynomials converges very fast to the exactly known spectrum for $N=1$. This of course implies even faster convergence rate for higher values of $N$.

Some applications of this result comes to mind. For example, in the context of the Wirtinger-Sobolev inequalities, which are the generalizations of the string ($N=1$) and rod ($N=2$) inequalities \cite{Finch}, one can benefit from the results derived here. A Wirtinger-Sobolev inequality is simply an inequality of the form
\begin{equation}
\int\limits_{-1}^1 {\left[ {U(x)} \right]^2 dx} \le C
\int\limits_{-1}^1 {U(x)\left[ {(-\Delta)^N U(x)} \right]dx}=C
\int\limits_{-1}^1 {\left[ U^{(N)}(x) \right]^2 dx}
 \label{eq:WSinqlty} \, ,
\end{equation}
where $U(x)$ is required to satisfy certain additional conditions, such as the BC above (\ref{eq:BC}). The variational formulation (\ref{eq:variational}) shows that the smallest eigenvalue $\lambda_0$ of the operator $\Delta^N$ indeed gives the best constant $C$, with $C=1/\lambda_0$. However, knowing the
whole spectrum of $\Delta^N$ gives much more information. For example, suppose that we know that a given function $U(x)$ is even (odd) and that it crosses zero $2j$ ($2j+1$) times. In that case, the best constants in the Wirtinger-Sobolev inequalities can be improved and become $C=1/\lambda_j$ (or $C=1/\lambda_{j+\frac{1}{2}}$ for the odd case). For large $N$, these quantities are determined using Eq.~(\ref{eq:lambdaN}), and for any $N$ they can be calculated efficiently using the numerical procedure described above. The key result here is that each zero crossing of $U(x)$ in the interval lowers the constant in the inequality (\ref{eq:WSinqlty}) by a factor proportional to $N$ (when $N$ is large).

Interesting generalizations of the present results include fractional values of $N$ \cite{Zoia,fractional}, other boundary conditions and more than one-dimensional Laplacian \cite{Pomeau}. Already for large, but not necessarily integer $N$, the results we derived for the eigenvalues hold even for the fractional case. Moreover, we believe that the approach presented here can be adapted to yield the whole spectrum of those operators in the large $N$ limit, and through a rapidly convergent numerical approach for any $N$. The main difficulty is to find a set of orthornormal polynomials for the case of fractional Laplacian which plays the same role as the associated Legendre polynomials for integer $N$'s.

A more challenging class of problems are those where the domain is not simple, such as a rectangle or a billiard. An approach in the spirit of the present one might allow some analytical insight through a $1/N$ expansion into problems that are traditionally heavily numerical. For example, for the $3D$ Laplacian case the problem can be posed as solving~\cite{Pomeau}
\begin{equation}
\left(\frac{d^2}{dr^2}-\frac{L(L+1)}{r^2}\right)^N(ru(r))=(-1)^N\lambda (ru(r))\,,
\end{equation}
with $u(r)$ satisfying the BC (\ref{eq:BC}). One can show that the appropriate set of orthonormal polynomials are
\begin{equation}
v_j(r)\propto r^L(1-r^2)^{N}P^{(2N,L+1/2)}_j(2 r^2-1)\,,
\end{equation}
where $P^{(\alpha,\beta)}_j(x)$ are the Jacobi polynomials. Using this observation, one can repeat the same procedure as presented here in order to obtain the whole set of eigenvalues.

\section*{Acknowledgements}

This work was supported by EEC PatForm Marie Curie action (E.K.). We thank Y. Pomeau for fruitful discussions. Laboratoire de Physique Statistique is associated with Universities Paris VI and Paris VII.

\appendix

\section{Derivation of the matrix elements $\hat \Delta _{m,j}^N$}

In this appendix we present how to obtain the matrix elements $\hat \Delta _{m,j}^N$ given in Eq.~(\ref{eq:matrix}). The starting point is the explicit series representation of the associated Legendre polynomials that can be obtained for example from Equation 8.812 in Ref.~\cite{G&R}
\begin{equation}
\fl P_{2N + 2j}^{2N} \left( x \right) = \frac{1}{{2^{2N + 2j}
}}\left( {1 - x^2 } \right)^N \sum\limits_{\ell  = 0}^j
{\frac{{\left( { - 1} \right)^{j - \ell } \left( {4N + 2j + 2\ell }
\right)!}}{{\left( {2\ell } \right)!\left( {j - \ell }
\right)!\left( {2N + j + \ell } \right)!}}x^{2\ell } } \
 \label{eq:ALP2} \, .
\end{equation}
We can now explicitly act on this representation with the operator
$\Delta^N$
\begin{eqnarray}
 \fl \Delta ^N P_{2N + 2j}^{2N} (x) = \frac{1}{{2^{2N + 2j} }}\sum\limits_{\ell  = 0}^j {\frac{{\left( { - 1} \right)^{j - \ell } \left( {4N + 2j + 2\ell } \right)!}}{{\left( {2\ell } \right)!\left( {j - \ell } \right)!\left( {2N + j + \ell } \right)!}}\Delta ^N \left[ {x^{2\ell } \left( {1 - x^2 } \right)^N } \right]}  =  \nonumber \\
 \fl \quad = \frac{1}{{2^{2N + 2j} }}\sum\limits_{\ell  = 0}^j {\frac{{\left( { - 1} \right)^{j - \ell } \left( {4N + 2j + 2\ell } \right)!}}{{\left( {2\ell } \right)!\left( {j - \ell } \right)!\left( {2N + j + \ell } \right)!}}\sum\limits_{m = 0}^{\min \left\{ {N,\ell } \right\}} {\frac{{N!\left( { - 1} \right)^{N - m} }}{{m!\left( {N - m} \right)!}}\frac{{\left( {2N + 2\ell  - 2m} \right)!}}{{\left( {2\ell  - 2m} \right)!}}x^{2\ell  - 2m} } }
 \label{eq:DNALP} \, .
\end{eqnarray}
This in turn can be written as $\Delta ^N P_{2N + 2j}^{2N} \left( x
\right) = \sum\limits_{i = 0}^j {b_i^{\left( {j,N} \right)} x^{2i}}$
with
\begin{equation}
\fl b_i^{\left( {j,N} \right)}  = \frac{{\left( { - 1} \right)^{N} \left( {2N + 2i}
\right)!}}{{2^{2N + 2j}\left( {2i} \right)!}}\sum\limits_{m = 0}^{j - i}
{{{{\left( { - 1} \right)^{j-i}\left( {4N + 2j + 2m + 2i} \right)!} \over {\left( {2m
+ 2i} \right)!\left( {j - m - i} \right)!\left( {2N + j + m + i}
\right)!}}}{{{N!} \over {m!\left( {N - m} \right)!}}}}
 \label{eq:bijN} \, .
\end{equation}
Using the integral 7.126-2 from Ref.~\cite{G&R}
\begin{equation}
\fl \int\limits_{ - 1}^1 {x^{2i} P_{2\ell }^{2N} \left( x \right)dx}
= {\textstyle{{\sqrt \pi  \Gamma \left( {i + {\textstyle{1 \over
2}}} \right)} \over {2^{4N} \Gamma \left( {N + {\textstyle{1 \over
2}}} \right)\Gamma \left( {N + i + {\textstyle{3 \over 2}}}
\right)}}}{\textstyle{{\left( {2\ell  + 2N} \right)!} \over {\left(
{2\ell  - 2N} \right)!}}}{}_3F_2 \left( {\begin{array}{*{20}c}
   {N + \ell  + {\textstyle{1 \over 2}},N - \ell ,N + 1}  \\
   {2N + 1,N + i + {\textstyle{3 \over 2}}}  \\
\end{array};1} \right)
 \label{eq:momentsALP} \, ,
\end{equation}
we can simply obtain Eq.~(\ref{eq:matrix}).

\section{Large $N$ asymptotic expansion of the matrix elements}

In this appendix we show how to obtain the large $N$ expansion of the matrix elements $\hat \Delta _{m,j}^N$ summarized in Eq.~(\ref{eq:deltajm}). In order to facilitate the calculation we assume from now on that $m \ge j$, since the matrix is symmetric and so the case $m<j$ is trivially obtained. The first step is to simplify the hypergeometric function ${}_3F_2$. We use formulas 7.527-1 and 3.259-3 from Ref.~\cite{G&R} to obtain the following integral representations
\begin{eqnarray}
 \fl
\begin{array}{l}
 {}_3F_2 \left( {\begin{array}{*{20}c}
   {2N + j + {\textstyle{1 \over 2}}, - j,N + 1}  \\
   {2N + 1,N + i + {\textstyle{3 \over 2}}}  \\
\end{array};1} \right) =  \\
 \quad \underbrace  = _{7.527 - 1}\frac{1}{{B\left( {N + 1,i + {\textstyle{1 \over 2}}} \right)}}\int\limits_0^\infty  {\left( {1 - e^{ - x} } \right)^{i - {\textstyle{1 \over 2}}} e^{ - \left( {N + 1} \right)x} {}_2F_1 \left( {\begin{array}{*{20}c}
   {2N + j + {\textstyle{1 \over 2}}, - j}  \\
   {2N + 1}  \\
\end{array};e^{ - x} } \right)dx}  =  \\
 \quad \underbrace  = _{3.259 - 3}\frac{{\int\limits_0^\infty  {\left( {1 - e^{ - x} } \right)^{i - {\textstyle{1 \over 2}}} e^{ - \left( {N + 1} \right)x} \int\limits_0^\infty  {y^{2N + j - {\textstyle{1 \over 2}}} \left( {1 + y} \right)^{ - \left( {2N + j + 1} \right)} \left[ {1 + \left( {1 - e^{ - x} } \right)y} \right]^j dy} dx} }}{{B\left( {N + 1,i + {\textstyle{1 \over 2}}} \right)B\left( {2N + j + {\textstyle{1 \over 2}},{\textstyle{1 \over 2}} - j} \right)}} \\
 \end{array}
 \label{eq:hyper1} \, ,
\end{eqnarray}
where $B(x,y)$ is the Beta function. In addition we expand $\left[ {1 + \left( {1 - e^{ - x} } \right)y} \right]^j  = \sum\limits_{\ell  = 0}^j {{\textstyle{{j!} \over {\ell !\left( {j - \ell } \right)!}}}\left( {1 - e^{ - x} } \right)^\ell y^\ell}$ and so
\begin{equation}
\fl {}_3F_2 \left( {\begin{array}{*{20}c}
   {2N + j + {\textstyle{1 \over 2}}, - j,N + 1}  \\
   {2N + 1,N + i + {\textstyle{3 \over 2}}}  \\
\end{array};1} \right) = {\textstyle{{\sum\limits_{\ell  = 0}^j {{\textstyle{{j!} \over {\ell !\left( {j - \ell } \right)!}}}} \left[ {\int\limits_0^\infty  {y^{2N + j + \ell  - {\textstyle{1 \over 2}}} \left( {1 + y} \right)^{ - \left( {2N + j + 1} \right)} dy} } \right]\left[ {\int\limits_0^\infty  {\left( {1 - e^{ - x} } \right)^{i + \ell  - {\textstyle{1 \over 2}}} e^{ - \left( {N + 1} \right)x} dx} } \right]} \over {B\left( {N + 1,i + {\textstyle{1 \over 2}}} \right)B\left( {2N + j + {\textstyle{1 \over 2}},{\textstyle{1 \over 2}} - j} \right)}}}
 \label{eq:hyper2} \, .
\end{equation}
After using $\int\limits_0^\infty  {y^a \left( {1 + y} \right)^b dy}
= {\textstyle{{\Gamma \left( {a + 1} \right)\Gamma \left( { - 1 - a
- b} \right)} \over {\Gamma \left( { - b} \right)}}}$ and $
\int\limits_0^\infty  {\left( {1 - e^{ - x} } \right)^a e^{ - bx}
dx}  = B\left( {a + 1,b} \right)$ we get
\begin{equation}
\fl {}_3F_2 \left( {\begin{array}{*{20}c}
   {2N + j + {\textstyle{1 \over 2}}, - j,N + 1}  \\
   {2N + 1,N + i + {\textstyle{3 \over 2}}}  \\
\end{array};1} \right) = \sum\limits_{\ell  = 0}^j {{\textstyle{{j!} \over {\ell !\left( {j - \ell } \right)!}}}{\textstyle{{B\left( {2N + j + \ell  + {\textstyle{1 \over 2}},{\textstyle{1 \over 2}} - \ell } \right)} \over {B\left( {2N + j + {\textstyle{1 \over 2}},{\textstyle{1 \over 2}} - j} \right)}}}{\textstyle{{B\left( {N + 1,i + \ell  + {\textstyle{1 \over 2}}} \right)} \over {B\left( {N + 1,i + {\textstyle{1 \over 2}}} \right)}}}}
 \label{eq:hyper3} \, .
\end{equation}
Last, for large $N$ we can expand the Beta functions and get to leading order (subleading orders are not difficult to obtain)
\begin{equation}
\fl {}_3F_2 \left( {\begin{array}{*{20}c}
   {2N + j + {\textstyle{1 \over 2}}, - j,N + 1}  \\
   {2N + 1,N + i + {\textstyle{3 \over 2}}}  \\
\end{array};1} \right) = \frac{{\sqrt \pi  {}_2F_1 \left( {\begin{array}{*{20}c}
   {i + {\textstyle{1 \over 2}}, - j}  \\
   {{1 \mathord{\left/
 {\vphantom {1 2}} \right.
 \kern-\nulldelimiterspace} 2}}  \\
\end{array};2} \right)}}{{2^j N^j \Gamma \left( {{\textstyle{1 \over 2}} - j} \right)}} + O\left( {\frac{1}{{N^{j + 1} }}} \right)
 \label{eq:hyper4} \, .
\end{equation}
Next, we expand the factorials in $b_i^{\left( {j,N} \right)}$ as defined in Eq.~(\ref{eq:bijN})
\begin{equation}
b_i^{\left( {j,N} \right)}  = \pi \frac{{\left( { - 1} \right)^{N +
j + i} 2^{2i + 2j + 6N + {\textstyle{3 \over 2}}} e^{ - 4N} N^{4N +
3j + i + {\textstyle{1 \over 2}}} }}{{\left( {2i} \right)!j!\left(
{j - i} \right)!\Gamma \left( {j + {\textstyle{1 \over 2}}}
\right)}}\left[ {1 + O\left( {\frac{1}{N}} \right)} \right]
 \label{eq:asymptb} \, ,
\end{equation}
so that
\begin{eqnarray}
 \sqrt {{\textstyle{{\left( {4N + 4m + 1} \right)\left( {4N + 4j + 1} \right)\left( {4N + 2m} \right)!\left( {2j} \right)!} \over {\left( {4N + 2j} \right)!\left( {2m} \right)!}}}} {\textstyle{{\sqrt \pi  } \over {2^{4N + 1} \Gamma \left( {N + {\textstyle{1 \over 2}}} \right)}}}{\textstyle{{\Gamma \left( {i + {\textstyle{1 \over 2}}} \right)} \over {\Gamma \left( {N + i + {\textstyle{3 \over 2}}} \right)}}}b_i^{\left( {j,N} \right)} \sim \nonumber \\
 \quad \sim \sqrt 2 \left( {2N} \right)!\sqrt {{\textstyle{{\left( {2j} \right)!} \over {\left( {2m} \right)!}}}} {\textstyle{{\left( { - 1} \right)^{N + j + i} 2^{2i + 2m} N^{2j + m} \Gamma \left( {i + {\textstyle{1 \over 2}}} \right)} \over {\left( {2i} \right)!j!\left( {j - i} \right)!\Gamma \left( {j + {\textstyle{1 \over 2}}} \right)}}}
 \label{eq:asymptb2} \, .
\end{eqnarray}
After combining all these results, and some simple additional algebra we obtain Eq.~(\ref{eq:deltajm}).

\end{document}